\documentclass[%
 reprint,
 amsmath,amssymb,
 aps,
 pra,
]{revtex4-1}

\usepackage{graphicx}
\usepackage{dcolumn}
\usepackage{bm}
\usepackage{float}
\usepackage{xcolor}
\usepackage{amsmath}
\usepackage{mathtools}
\usepackage{etoolbox}
\usepackage{physics}
\usepackage{dsfont}
\apptocmd{\thebibliography}{\raggedright}{}{}
\usepackage{amsmath}
\usepackage[titletoc]{appendix}

\usepackage{soul}
\usepackage{alphabeta}

\def \beq {\begin{equation}}
\def \eeq {\end{equation}}

\begin{document}

\title {The non-Hermitian landscape of autoionization} 

\author{G. Mouloudakis$^{1,2}$}
 \email{gmouloudakis@physics.uoc.gr}

\author{P. Lambropoulos$^{1,2}$}%

\affiliation{${^1}$Department of Physics, University of Crete, P.O. Box 2208, GR-71003 Heraklion, Crete, Greece
\\
${^2}$Institute of Electronic Structure and Laser, FORTH, P.O.Box 1527, GR-71110 Heraklion, Greece}

\date{\today}

\begin{abstract}
We report on the existence of exceptional points (EPs) in single-resonance autoionization and provide analytical expressions for their positions in parameter space, in terms of the Fano asymmetry parameter. We additionally propose a reliable method for the experimental determination of EPs, based solely on information about their ionization probability as a function of the system parameters. The links between EPs, the maxima of the asymmetric profile and the effective decay rate of the ground state are investigated in detail. Quantitative numerical examples pertaining to the doubly excited $2s2p({}^1P)$ state of Helium confirm the validity of our formulation and results. In addition to unveiling hidden aspects of autoionization, our treatment and results provide a benchmark for the exploration of EPs and their properties in a variety of materials exhibiting Fano profiles with a broad perspective of possible applications.
\end{abstract}

\maketitle

\section{Introduction}
Autoionization (AI) belongs to a broad class of quantum phenomena involving discrete states (resonances) embedded in continua into which they decay. Examples, among others, are the  Breit-Wigner resonance in nuclear physics \cite{ref1}, in particle physics \cite{ref2,ref3}, in photonics \cite{ref4}  and of course in atoms and molecules \cite{ref5,ref6}, where the continuum is ionization or even dissociation; hence the term autoionization.  The literature on autoionization spans a vast range of topics, including the time-dependent formation of the autoionization profile \cite{ref7,ref8,ref9,ref10}, strong driving of autoionizing resonances (ARs) \cite{ref11,ref12,ref13,ref14,ref15,ref16,ref17}, the dynamics of doubly-resonant autoionization \cite{ref18,ref19}, and the effects of phase \cite{ref20,ref21} and statistical fluctuations \cite{ref22,ref23,ref24,ref25} of the laser field on the process.

ARs can be excited by radiation absorption or collisions and are infinite in number, with the spacing between them decreasing with increasing excitation energy. Yet, there are cases in which one or more resonances are  separated in energy by significantly more than their width, qualifying as isolated resonances, with the doubly excited $2s2p({}^1P)$ state of Helium being the prototype of an isolated AR, which continues revealing novel aspects, as attested by the ongoing streams of papers to this day \cite{ref13,ref14,ref15,ref16,ref17}. It is in addition a perfect example of an open quantum system, with its dynamics governed by a non-Hermitian  effective Hamiltonian. As such it can serve as a benchmark for a broad class of non-Hermitian systems in a variety of materials, exhibiting the same profile with possibilities of significant technological impact \cite{ref4,ref26,ref27,ref28,ref29}. 

Non-Hermitian physics and its connection to parity-time ($\mathcal{P} \mathcal{T}$) symmetry, was introduced as an axiomatic theory in the seminal papers of C. Bender et al. \cite{ref30,ref31,ref32,ref33,ref34}. Soon thereafter, it was pointed out that effective Hamiltonians describing the dynamics of open quantum systems, inevitably are non-Hermitian \cite{ref35}. The boundary between the unbroken and broken $\mathcal{P} \mathcal{T}$ symmetry of such Hamiltonians \cite{ref36,ref37} is marked by the presence of exceptional points (EPs) \cite{ref38,ref39,ref40,ref41}, i.e., points in the parameter space where two or more eigenvalues coalesce, while their corresponding eigenvectors become parallel. Tracking the positions of these points in the parameter space of an open quantum system is crucial, as they provide insight into the range of parameters where the system undergoes abrupt phase transitions \cite{ref42} and enhanced sensitivity \cite{ref43,ref44,ref45,ref46,ref47}. Several approaches for understanding phenomena related to quasi-bound states embedded in continua using complex spectral analysis have been presented in the past, applied to various systems such as two-channel quantum wires \cite{ref48,ref26}, semi-infinite superlattices with embedded impurities \cite{ref27}, discrete states coupled to continua containing Van Hove singularities at their threshold \cite{ref28}, as well as systems involving laser-induced population trapping via strong coupling of ARs in atoms \cite{ref49}.

In this paper, we employ the powerful analysis of EPs in order to unveil hidden aspects of ARs. Focusing on the conditions for encountering EPs in single-resonance autoionization,  we derive analytical expressions revealing their positions in parameter space. Moreover, we show how the amount of ionization of the atom, which can be determined experimentally, contains information about the positions of EPs,  documented by numerical examples for the  $2s2p({}^1P)$ state of Helium. Finally, we demonstrate the connection between the presence of EPs, the maxima of the typical asymmetric profile of autoionization and the effective decay rate of the atomic ground state.

\section{Theoretical Background}

Our system consists of a ground state $\ket{g}$ coupled to an isolated quasi-bound resonance $\ket{a}$ through a linearly polarized field with frequency $\omega$, as well as a continuum of states denoted by $\ket{E}$, coupled both to $\ket{g}$ and $\ket{a}$. The field that drives the $\ket{g} \longleftrightarrow \ket{a}$ and $\ket{g} \longleftrightarrow \ket{E}$ transitions is of the form
$E(t)= \frac{1}{2} \left[ \mathcal{E} e^{-i ω t} + \mathcal{E}^* e^{i ω t} \right]$.

The Hamiltonian of the system $\hat{\mathcal{H}} = \hat{\mathcal{H}}_0 + \hat{\mathcal{V}} + \hat{\mathcal{D}}$ consists of three parts; namely, the free-atom Hamiltonian $\hat{\mathcal{H}}_0$ with $\hat{\mathcal{H}}_0 \ket{j} = \omega_j \ket{j}$, $j=g,a,E$, the configuration interaction $\hat{\mathcal{V}}$ that couples the discrete autoionizing resonance to the continuum, as well as the dipole interaction $\hat{\mathcal{D}}$ that couples both $\ket{g}$ and $\ket{a}$ to the continuum, given by the relation $\hat{\mathcal{D}}= \wp E(t)$, where $\wp = \vec{\wp} \cdot \hat{e}$ is the projection of the electric dipole moment operator on the polarization direction of the electric field amplitude, denoted by $\hat{e}$. Note that throughout our calculations we set $\hbar=1$.

The wavefunction of the system at times $t \geq 0$ is given by:
\beq
\ket{\psi(t)}= c_g(t) \ket{g} + c_a(t) \ket{a} + \int dE c_E(t) \ket{E}.
\label{Wavefunctiom}
\eeq
The time-dependent Schr\"{o}dinger equation (TDSE) in view of Eq. \eqref{Wavefunctiom} reduce to the following set of equations:

\begin{subequations}
\beq
i \partial_t c_g(t) = ω_g c_g(t) + \mathcal{D}_{ga} c_a(t) + \int dE \mathcal{D}_{gE} c_E(t),
\eeq

\beq
i \partial_t c_a(t) = ω_a c_a(t) + \mathcal{D}_{ga}^* c_g(t) + \int dE \mathcal{V}_{aE} c_E(t),
\eeq

\beq
i \partial_t c_E(t) = ω_E c_E(t) + \mathcal{D}_{gE}^* c_g(t) + \mathcal{V}_{aE}^* c_a(t),
\eeq
\end{subequations}
where we adopted the notation $\partial_t \equiv \frac{\partial}{\partial t}$. Introducing the slowly varying amplitudes according to the transformations $\tilde{c}_g(t) = c_g(t) e^{i \omega_g t}$, $\tilde{c}_a(t) = c_a(t) e^{i (\omega_g + \omega) t}$ and $\tilde{c}_E(t) = c_E(t) e^{i (\omega_g + \omega ) t}$, the above set of equations become:

\begin{subequations}
\beq
i \partial_t \Tilde{c}_g(t) = \mathcal{D}_{ga} e^{-i \omega t} \tilde{c}_a(t)  + \int dE \mathcal{D}_{gE} e^{-i \omega t} \tilde{c}_E(t), 
\label{c_g Eqn}
\eeq

\beq
i \partial_t \tilde{c}_a(t) = (\omega_a - \omega_g - \omega) \tilde{c}_a(t) + \mathcal{D}_{ga}^* e^{i \omega t} \tilde{c}_g(t)  + \int dE \mathcal{V}_{aE} \tilde{c}_E(t),
\label{c_a Eqn}
\eeq

\beq
i \partial_t \tilde{c}_E(t) = ( \omega_E- \omega_g - \omega ) \tilde{c}_E(t) + \mathcal{D}_{gE}^* e^{i \omega t} \tilde{c}_g(t)  + \mathcal{V}_{aE}^* \tilde{c}_a(t).
\label{c_E Eqn}
\eeq
\end{subequations}

We now eliminate the continuum adiabatically by setting $\partial_t \tilde{c}_E(t)$=0, treating it as a sink. Under this assumption Eq. \eqref{c_E Eqn} leads to:
\beq
\tilde{c}_E(t)=\frac{\mathcal{D}_{gE}^* e^{i \omega t}}{\omega_g + \omega - \omega_E} \tilde{c}_g(t) + \frac{\mathcal{V}_{aE}^*}{\omega_g + \omega - \omega_E} \tilde{c}_a(t).
\label{c_E after elimination}
\eeq
Substitution of Eq. \eqref{c_E after elimination} back to Eqs. \eqref{c_g Eqn} and \eqref{c_a Eqn}, yields:

\begin{subequations}
\beq
\begin{split}
i \partial_t \Tilde{c}_g(t) & =  \mathcal{D}_{ga} e^{-i \omega t} \tilde{c}_a(t)  + \int{ dE \frac{\abs{\mathcal{D}_{gE}}^2}{\omega_g + \omega - \omega_E} \tilde{c}_g(t)}  \\
& + \int{ dE \frac{ \mathcal{V}_{aE}^* \mathcal{D}_{gE} e^{-i \omega t}}{\omega_g + \omega - \omega_E} \tilde{c}_a(t)},
\end{split}
\eeq

\beq
\begin{split}
i \partial_t \tilde{c}_a(t) & =  (\omega_a - \omega_g - \omega) \tilde{c}_a(t) + \int{ dE \frac{ \mathcal{V}_{aE} \mathcal{D}_{gE}^* e^{i \omega t}}{\omega_g + \omega - \omega_E} \tilde{c}_g(t)}   \\
& + \mathcal{D}_{ga}^* e^{i \omega t} \tilde{c}_g(t)  + \int{ dE \frac{\abs{\mathcal{V}_{aE}}^2}{\omega_g + \omega - \omega_E} \tilde{c}_a(t)}.
\end{split}
\eeq
\end{subequations}
By substituting the matrix elements $\mathcal{D}_{ga}=\wp_{ga} E(t)$ and $\mathcal{D}_{gE}=\wp_{gE} E(t)$ of the dipole interaction operator in the above set of equations and adopting the rotating-wave approximation (RWA) which implies the neglection of the fast-oscillating anti-resonant time-dependent exponentials, we obtain:
\begin{subequations}
\beq
\begin{split}
i \partial_t \Tilde{c}_g(t) & =  \int{ dE \frac{\abs{\Omega_{gE}}^2}{\omega_g + \omega - \omega_E} \tilde{c}_g(t)}  \\
& + \left( \Omega_{ga} + \int{ dE \frac{ \mathcal{V}_{aE}^* \Omega_{gE}}{\omega_g + \omega - \omega_E}} \right) \tilde{c}_a(t),
\end{split}
\eeq

\beq
\begin{split}
i \partial_t \tilde{c}_a(t) & =  \left( \Omega_{ga}^* + \int{ dE \frac{ \mathcal{V}_{aE} \Omega_{gE}^*}{\omega_g + \omega - \omega_E}} \right) \tilde{c}_g(t) \\
& + \left[ (\omega_a - \omega_g - \omega) + \int{ dE \frac{\abs{\mathcal{V}_{aE}}^2}{\omega_g + \omega - \omega_E}} \right] \tilde{c}_a(t), 
\end{split}
\eeq
\label{Semi-final set}
\end{subequations}
where we have introduced the definitions $\Omega_{ga} \equiv \frac{1}{2} \wp_{ga} \mathcal{E}^*$ and $\Omega_{gE} \equiv \frac{1}{2} \wp_{gE} \mathcal{E}^*$.
The above set of equations is simplified considerably by using  the identity
\beq
\begin{split}
\lim_{η \to 0^+} \frac{1}{\omega_g + \omega - \omega_E +i η} & = \mathds{P}\frac{1}{\omega_g + \omega - \omega_E} \\
& - i \pi δ(\omega_g + \omega - \omega_E),
\end{split}
\label{lim_identity}
\eeq
where $\mathds{P}$ denotes the principal value part and $δ(x)$ is the Dirac delta function. In view of Eq. \eqref{lim_identity}, Eqs. \eqref{Semi-final set} can ultimately be written as:
\begin{subequations}
\beq
i \partial_t \Tilde{c}_g(t) = \left( S_g - i \frac{\gamma}{2} \right) \Tilde{c}_g(t) + \tilde{\Omega}_{ga} \left( 1 - \frac{i}{q} \right) \Tilde{c}_a(t),
\eeq

\beq
i \partial_t \Tilde{c}_a(t) = \tilde{\Omega}_{ag}\left( 1 - \frac{i}{q} \right) \Tilde{c}_g(t) - \left( Δ + i \frac{Γ}{2} \right) \Tilde{c}_a(t),
\eeq
\label{Final_set}
\end{subequations}
where $S_g \equiv \mathds{P} \int{ dE \frac{\abs{\Omega_{gE}}^2}{\omega_g + \omega - \omega_E}}$ and $ \left. \gamma \equiv 2 \pi \abs{\Omega_{gE}}^2 \right\vert_{\omega_E =\omega_g + \omega}$ are, respectively, the light-induced shift and the ionization rate of the ground state, whereas $F_a \equiv \mathds{P} \int{ dE \frac{\abs{\mathcal{V}_{aE}}^2}{\omega_g + \omega - \omega_E}}$ and $ \left. \Gamma \equiv 2 \pi \abs{\mathcal{V}_{aE}}^2 \right\vert_{\omega_E =\omega_g + \omega}$ are, respectively, the self-energy shift and the autoionization rate of the state $\ket{a}$. Moreover, $\tilde{\Omega}_{ga} \equiv \Omega_{ga} + \mathds{P} \int{ dE \frac{ \mathcal{V}_{aE}^* \Omega_{gE}}{\omega_g + \omega - \omega_E}}$ is the generalized Rabi frequency of the $\ket{g} \longleftrightarrow \ket{a}$ transition, $q \equiv \frac{\tilde{\Omega}_{ga}}{\pi \Omega_{gE} \mathcal{V}_{aE}^*}$ is the Fano asymmetry parameter and $\Delta \equiv \omega - (\omega_a - F_a -\omega_g)$ is the detuning between the frequency of the driving field and the frequency of the $\ket{g} \longleftrightarrow \ket{a}$ transition, including the self-energy shift of state $\ket{a}$.

\begin{figure}[t]
	\centering
	\includegraphics[width=6.0cm]{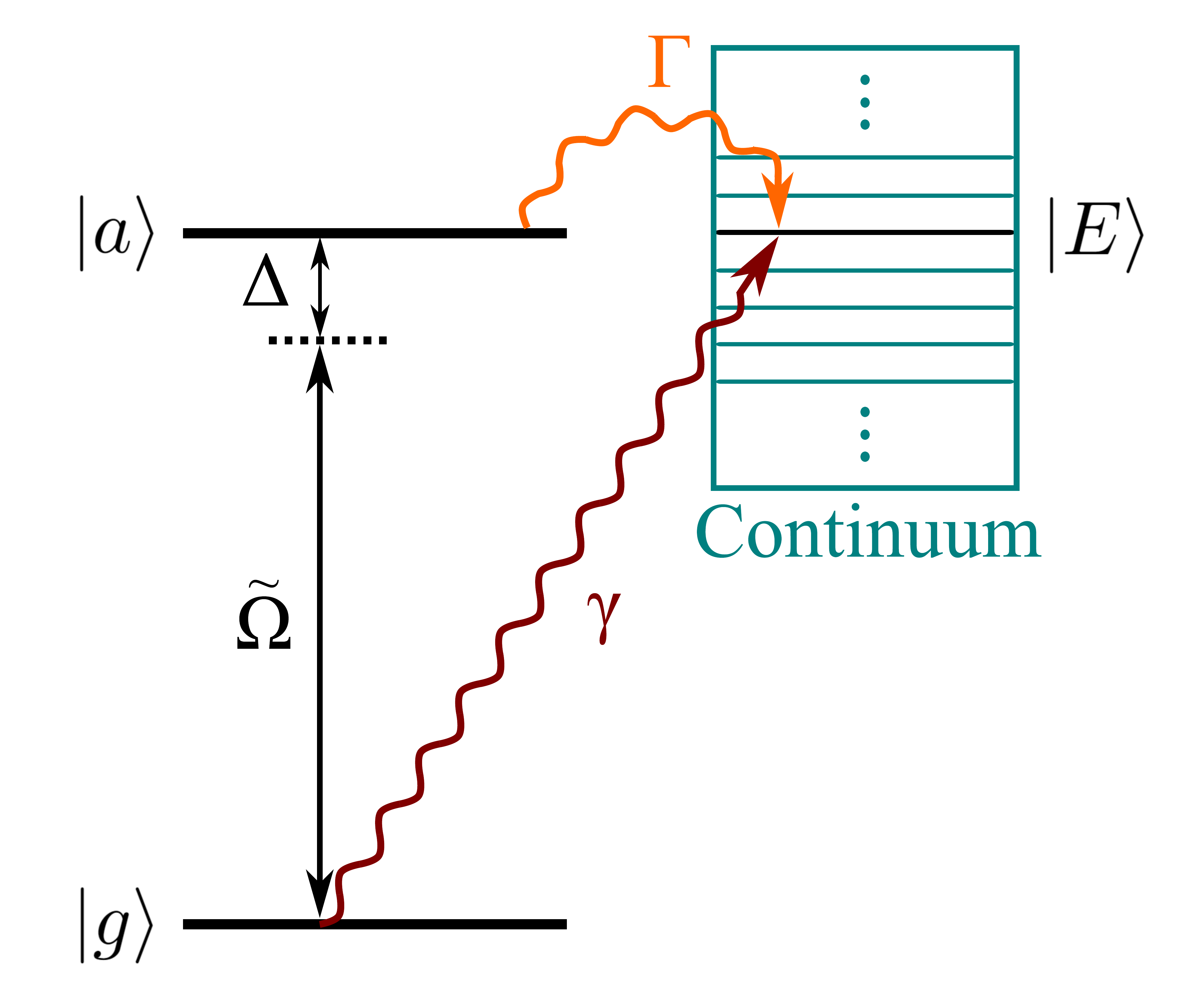}
		\caption[Fig1]{Schematic representation of the system at study. The ground state $\ket{g}$ of an atom that is ionized with a rate $\gamma$, is coupled to an AR $\ket{a}$ via a linearly polarized field that drives the $\ket{g} \longleftrightarrow \ket{a}$ transition with a generalized Rabi frequency $\tilde{\Omega}$. The frequency of the driving field is detuned by $\Delta$ from the energy separation of the two states and the AR decays into the continuum with an autoionization rate $\Gamma$.}
		\label{Fig1}
\end{figure}

The set of Eqs. \eqref{Final_set} can be written as:
\beq
i \partial_t 
\begin{bmatrix}
\Tilde{c}_g(t) \\
\Tilde{c}_a(t)
\end{bmatrix}
= \hat{\mathcal{H}}_{\text{eff}}
\begin{bmatrix}
\Tilde{c}_g(t) \\
\Tilde{c}_a(t)
\end{bmatrix},
\eeq
with

\beq
\hat{\mathcal{H}}_{\text{eff}} \equiv 
\begin{bmatrix}
S_g -i \frac{\gamma}{2} & \tilde{\Omega} \left( 1-\frac{i}{q} \right)  \\
\tilde{\Omega} \left( 1-\frac{i}{q} \right)  &   - \Delta -i \frac{\Gamma}{2}
\end{bmatrix},
\label{Effective_Hamiltonian}
\eeq
where in order to simplify notation we have introduced the definition $\tilde{\Omega} \equiv \tilde{\Omega}_{ga}$. $\hat{\mathcal{H}}_{\text{eff}}$ is the effective Hamiltonian governing the dynamics of our system under the adiabatic elimination of the continuum and  RWA approximations. Combining the definitions of $\gamma$, $\Gamma$ and $\tilde{\Omega}$ given above, we obtain  $q^2=4 \tilde{\Omega}^2/(\gamma\Gamma)$ which is independent of intensity and provides a very useful relation between the parameters of the effective Hamiltonian. A schematic representation of our system in terms of these parameters is depicted in Fig. 1.

The effective Hamiltonian of Eq. \eqref{Effective_Hamiltonian} is obviously non-Hermitian, not only due to the presence of the diagonal decay terms in the energies of the ground state and $\ket{a}$, but also due to the presence of non-zero imaginary parts in the off-diagonal terms reflecting the driving of the $\ket{g} \longleftrightarrow \ket{a}$ transition. Diagonalization of $\hat{\mathcal{H}}_{\text{eff}}$ leads to the following set of eigenvalues:

\beq
\begin{split}
\lambda_{1,2} & = - \frac{1}{2} \left[ \Delta + i  \frac{\left(\gamma+\Gamma\right)}{2} \right] \\
& \pm \frac{1}{4} \sqrt{16 \left( 1-\frac{i}{q} \right)^2 \tilde{\Omega}^2 - \left( \gamma - \Gamma +2i \Delta \right)^2}.
\end{split}
\eeq
At first sight, owing to the presence of imaginary parts in the radicands, the spectra of $\hat{\mathcal{H}}_{\text{eff}}$ appear not to exhibit EPs. However, if the detuning is set to  $\Delta=\Delta^s \equiv 2q \gamma \Gamma / (\Gamma - \gamma)$, $\gamma \neq \Gamma$ and we eliminate $\gamma$ via the relation $\gamma = 4 \Omega^2 /(q^2 \Gamma)$, we obtain:

\begin{widetext}
\beq
\lambda_{1,2}=  \frac{4 \tilde{\Omega}^2 q \Gamma}{4 \tilde{\Omega}^2 - q^2  \Gamma^2}  - i \left(  \frac{ \tilde{\Omega}^2}{q^2 \Gamma} + \Gamma/4 \right)  \pm \frac{1}{4 q \abs{q} \Gamma} \sqrt{ - {\left( \frac{ 4 \tilde{\Omega}^2 + q^2 \Gamma^2}{ 4 \tilde{\Omega}^2 - q^2 \Gamma^2 } \right)}^2 \left[  16 \tilde{\Omega}^4 - 8 \tilde{\Omega}^2 \Gamma^2 q^2 \left( 1 + 2 q^2 \right) + q^4 \Gamma^4 \right]  }, \: \tilde{\Omega} \neq \frac{\abs{q}Γ}{2}
\label{Eigenvalues}
\eeq
\end{widetext}

Observe now that choosing $\Delta$ = $\Delta^s$ results to a set of eigenvalues with real radicands. Note that Eq. \eqref{Eigenvalues} holds for $ \tilde{\Omega} \neq \abs{q}Γ/2$ which is equivalent to $\gamma \neq \Gamma$. For $ \tilde{\Omega} = \abs{q}Γ/2$, i.e. $\gamma=\Gamma$, the radicand is complex for every value of $\Delta$. The details of the physical significance of $\Delta^s$ for our system will become clear later. We should also note that the value of $\Delta^s$  resulting to real radicands depend on the intensity of the driving field, which in turn determines the value of $\tilde{\Omega}$. The relation between $\Delta^s$ and $\tilde{\Omega}$ is $  \frac{\Delta^s ( \tilde{\Omega} )}{Γ} = \frac{8 q { \left( \frac{\tilde{\Omega}}{Γ} \right)}^2}{q^2 - 4 { \left( \frac{\tilde{\Omega}}{Γ} \right)}^2} $, $\tilde{\Omega} \neq \abs{q}Γ/2$, which results upon substitution of $\gamma = 4 \Omega^2 / (q^2 \Gamma)$ in the expression $\Delta^s \equiv 2q\gamma \Gamma / (\Gamma - \gamma)$, $\gamma \neq \Gamma$.

We are interested in the values of the coupling $\tilde{\Omega}$ that nullify the radicands of Eq. \eqref{Eigenvalues}. The radicands become zero when
\beq
 16 \tilde{\Omega}^4 - 8 \tilde{\Omega}^2 \Gamma^2 q^2 \left( 1 + 2 q^2 \right) + q^4 \Gamma^4=0,
\eeq
and the positive roots of the above equation are
\beq
 \frac{\tilde{\Omega}_{\pm}}{Γ} = \frac{1}{2} \left( \abs{q} \sqrt{1 + q^2} \pm q^2 \right).
\label{EPs}
\eeq
It is easy to verify that for both $\tilde{\Omega}=\tilde{Ω}_{+}$ and $\tilde{\Omega}=\tilde{Ω}_{-}$, given that $\Delta=\Delta^s$, the eigenvectors of $\hat{\mathcal{H}}_{\text{eff}}$ coalesce, respectively, to the states $\ket{\psi_{+}}=\left( - i \ket{g} + \ket{a} \right)/ \sqrt{2}$ and $\ket{\psi_{-}}=\left( i \ket{g} + \ket{a} \right)/ \sqrt{2}$. Therefore the points $( \tilde{Ω}_{\pm}, \Delta^s_{\pm} )$ in parameter space, where $\Delta^s_{\pm} \equiv \Delta^s ( \tilde{Ω}_{\pm} )$, are EPs of $\hat{\mathcal{H}}_{\text{eff}}$.

\begin{figure}[t] 
	\centering
	\includegraphics[width=8.5cm]{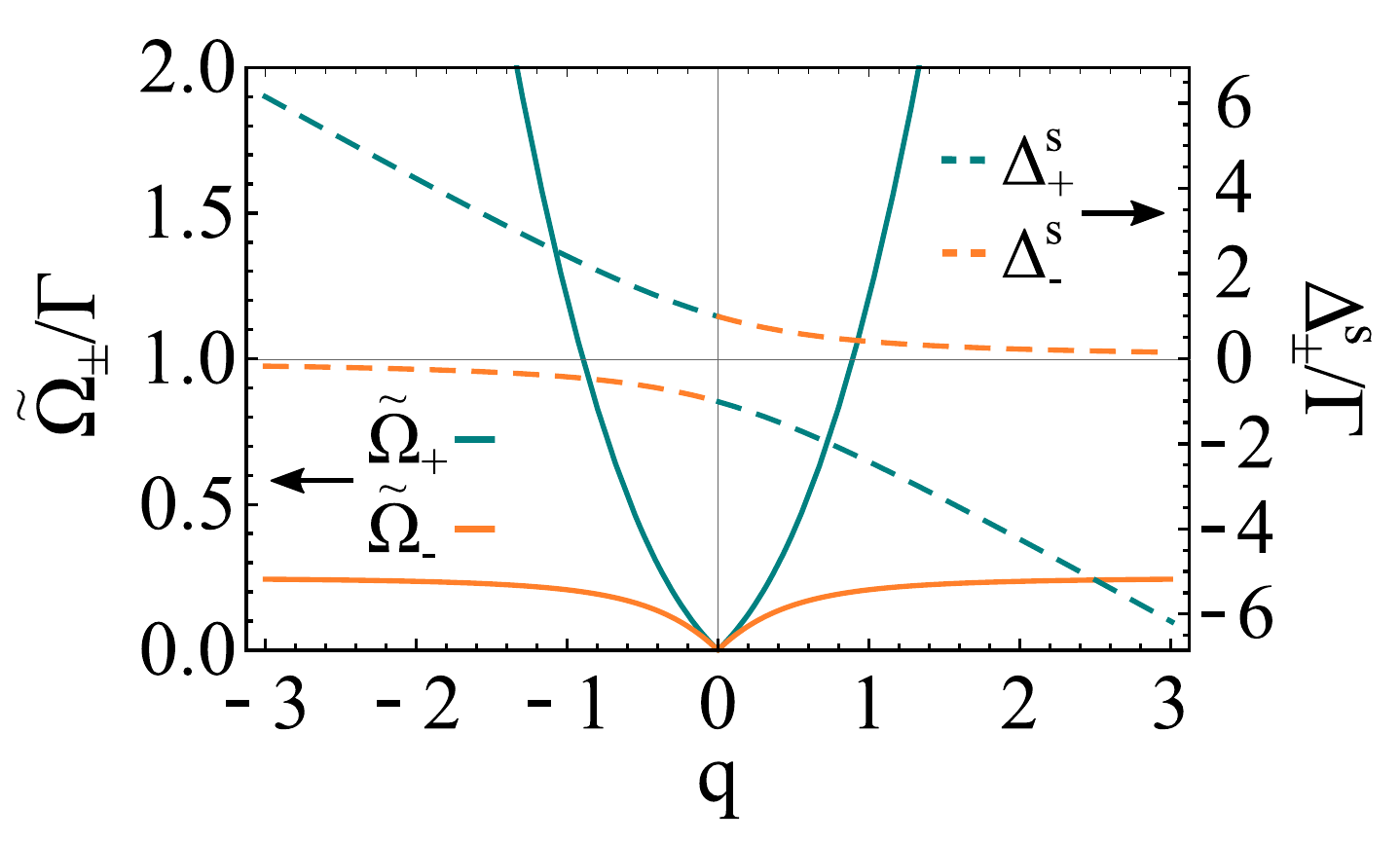}
		\caption[Fig2]{Dependence of the exceptional points $( \tilde{Ω}_{\pm}, \Delta^s_{\pm} )$ on the asymmetry parameter $q$. Solid teal line: $\Tilde{\Omega}_{+}$, dashed teal line: $\Delta^s_{+}$, solid orange line: $\Tilde{\Omega}_{-}$ and dashed orange line: $\Delta^s_{-}$. For each value of q there exist two exceptional points.}
		\label{Fig2}
\end{figure}

\begin{figure*}
	\centering
	\includegraphics[width=13.5cm]{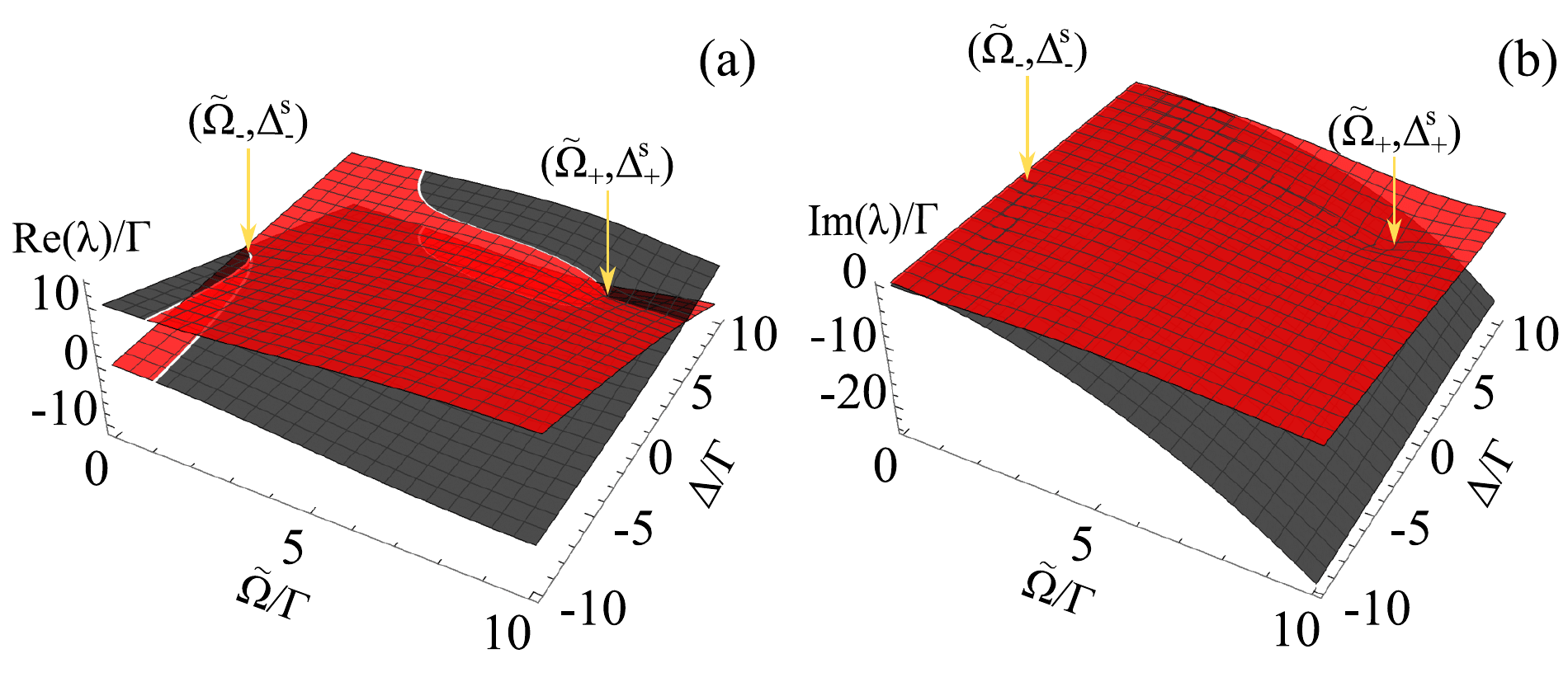}
		\caption[Fig3]{(a) Real and (b) imaginary parts of the eigenvalues $λ_1$ (red surface) and $λ_2$ (black surface) a function of the parameters $\Tilde{\Omega}$ and $Δ$, for $q=-2.79$. The yellow arrows mark the positions of the exceptional points at $( \Tilde{\Omega}, Δ) = ( \Tilde{\Omega}_{-}, \Delta^s_{-} ) = ( 0.2424Γ, -0.1738Γ )$ and $( \Tilde{\Omega}, Δ) = ( \tilde{\Omega}_{+}, \Delta^s_{+} ) = ( 8.0265Γ, 5.7538Γ )$, where the real and imaginary parts of the eigenvalues coalesce.}
		\label{Fig3}
\end{figure*}

\section{Results \& Discussion}

Interestingly, the EPs of the system measured in units of the autoionization width $\Gamma$, depend solely on the asymmetry parameter $q$, and there are two for any given value of the latter (Fig. 2). The interchange between $\Delta^s_{+}$ and $\Delta^s_{-}$ around $q=0$ is attributed to the presence of the absolute value of $q$ in the expressions of $\tilde{\Omega}_{\pm}$ and, subsequently, the appearance of $\abs{q}$ in the expressions of $\Delta^s_{\pm}$. It is important to note that the value of $q$ for a given AR is fixed, as it depends solely upon the corresponding matrix elements of the transitions involved in the process. In particular, for the process involving the driving of the $1s^2 ({}^1S) \longleftrightarrow 2s2p({}^1P)$ transition in Helium and the associated autoionization of the $2s2p({}^1P)$ AR, it is well established that $q \approx -2.79$ \cite{ref23,ref51}. 

Focusing hereafter on that isolated AR, we note that for $q = -2.79$, according to Eq. \eqref{EPs} and the relation between $Δ^s$ and $\Tilde{\Omega}$, the theory indicates the existence of two EPs at the positions $( \Tilde{\Omega}_{-}, \Delta^s_{-} ) = ( 0.2424Γ, -0.1738Γ )$ and $( \tilde{\Omega}_{+}, \Delta^s_{+} ) = ( 8.0265Γ, 5.7538Γ )$ in parameter space. In Fig. 3 we plot the real and imaginary parts of the eigenvalues as a function of $\tilde{\Omega}$ and $\Delta$ for $q = -2.79$ and indeed confirm the coalescence of the eigenvalues at the above positions in parameter space.

As noted above, tuning $\Delta$ to $\Delta^s$ is essential in order to ensure that the radicands appearing in the expressions of the eigenvalues become real. We can get a glimpse on the physical significance of $\Delta^s$ in the vicinity of an EP, by solving the time-dependent Schr\"{o}dinger equation using the effective Hamiltonian $\hat{\mathcal{H}}_{\text{eff}}$, and plotting the ionization probability of the atom $ ( P(t) = 1 - \abs{c_g(t)}^2 - \abs{c_a(t)}^2 )$ as a function of the detuning for $\Tilde{\Omega}=\Tilde{\Omega}_{-}$ (Fig. 4). Note that the ionization probability is calculated on $t=T$, where $T$ is the interaction time between the atom and the driving field. This calculation implies what is referred to as a square pulse, in which case the radiation is turned on at $t=0$ and off at the end of the pulse of duration $T$. In reality, however, the atom is exposed to a pulse of different shape, either because the source is pulsed or because the atom enters and exits a beam of radiation. It is, however, known that as long as the duration of the pulse, which means its full width at half maximum, is much longer (say by a factor of 5 or more) than the lifetime of the resonance, the results obtained through a square pulse, in the extreme case, would differ approximately by a factor of $\sqrt{2}$ \cite{ref23}. The lifetime of the $2s2p({}^1P)$ AR is about $17 fs$, whereas the development and observation of the EPs requires a pulse of much longer duration. This places no stringent demand on the source, since exposure times as long as $100 fs$ or more are routinely achievable. This equivalence is due to the fact that, the transients, caused by the sudden turn on and off of the square pulse, are overshadowed by the signal due to the flat part of the pulse, provided that the pulse duration is sufficiently long. Needless to add that long implies about 100 field cycles, in addition to the relation to the lifetime of the resonance, mentioned above. 

As expected, the ionization profile in Fig. 4 is asymmetric, transforming gradually to a "window" profile for sufficiently large interaction times, a phenomenon labelled "time saturation" in \cite{ref11}, reconfirmed most recently in \cite{ref15}. Interestingly, the position of the maximum of the asymmetric profile, denoted by $\Delta_m$, which is initially increasing as $T$ increases, eventually stabilizes at $\Delta^s_{-}$, as shown in the inset of Fig. 4. Therefore, for $\Tilde{\Omega}=\Tilde{\Omega}_{-}$, $\Delta^s ( \Tilde{\Omega}_{-} ) \equiv \Delta^s_{-}$ is the detuning which maximizes the ionization probability (to unity) for sufficiently large interaction times, which for the field intensity considered, translates to $T \approx 20Γ^{-1}$ or larger. It is important to note that this occurs only by tuning the parameters of the system to the exceptional point $( \Tilde{\Omega}_{-}, \Delta^s_{-} )$. For example, if we choose an intensity such that $\Tilde{\Omega}=0.1\Tilde{\Omega}_{-}$, the position of the maximum of the asymmetric profile stabilizes to $\Delta_m \approx -0.195 Γ$, whereas $\Delta^s ( 0.1 \Tilde{\Omega}_{-} ) = -0.0016 Γ$.

Although in most cases, the EPs of a system can be explored theoretically through diagonalization of the relevant effective Hamiltonian, the experimental determination of EPs most often is quite a challenging task, since in general the eigenenergies of a Hamiltonian are not amenable experimentally. Therefore one needs to identify EPs indirectly by studying their footprints on system observables. To that end, we employ a quantity widely used in the context of the Quantum Zeno effect in open quantum systems, namely, the effective decay rate of a state \cite{ref52}, defined as $\Gamma^j_{\text{eff}} (t) \equiv - \frac{1}{t} \ln [P_j (t)]$, $j=g,a$, where $P_j (t) = {\abs{c_j (t)}}^2$ is the population of state $\ket{j}$, $j=g,a$. The effective decay rate provides information about how the couplings between a given state and a set of other states or a continuum, modify the time evolution of that state's population. It turns out that the effective decay rate of the ground state, which can be readily determined experimentally, is remarkably sensitive to the EPs of our system, pinpointing their positions in parameter space. 

\begin{figure}[b] 
	\centering
	\includegraphics[width=7.8cm]{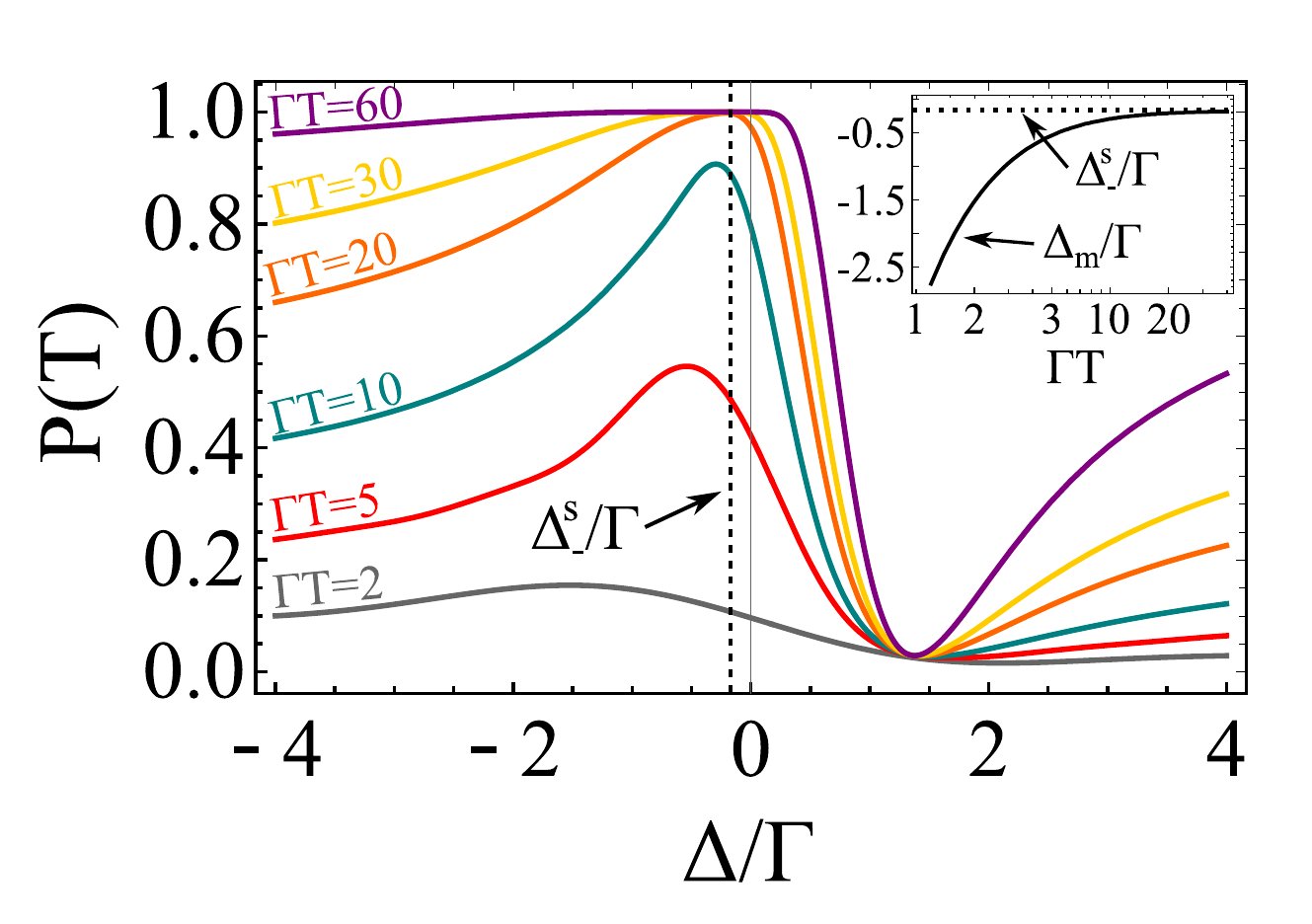}
		\caption[Fig4]{Ionization probability as a function of $\Delta$ for various interaction times $T$, $q=-2.79$ and $\tilde{\Omega}=\tilde{\Omega}_{-}=0.2424Γ$. The vertical dashed line marks the position of the detuning $\Delta^s_{-}=-0.1738 Γ$. Inset: Position of the peak of the asymmetric profile ($\Delta_m$) as a function of the interaction time $T$ (logarithmic scale) for $q=-2.79$ and $\tilde{\Omega}=\tilde{\Omega}_{-}$. The horizontal dotted line marks the position of the detuning $\Delta^s_-$.}
		\label{Fig4}
\end{figure}

In Fig. 5(a) we plot the effective decay rate of the ground state as a function of $\Tilde{\Omega}$ for $\Delta= \Delta^s ( \Tilde{\Omega} )$, which implies setting each time the detuning to a different value, depending on the  value of $\Tilde{\Omega}$ considered. Note that the effective decay rate is calculated at an interaction time $t=T$, which should be sufficiently large for the rate to be no longer modified with further increase of $T$. For $q=-2.79$, the effective decay rate is stabilized for $T \approx 20Γ^{-1}$ or larger, which is the same time scale as the one discussed in the results of Fig. 4. At such time scales it is easy to show that the population of $\ket{a}$ is practically negligible. Therefore the effective decay rate of the ground state is directly related to the measurable ionization probability $P(t)$, because $\Gamma^g_{\text{eff}} (t) \equiv - \frac{1}{t} \ln [P_g (t)] \cong \frac{1}{t} \ln [ 1 - P (t)]$. Clearly, the effective decay rate of the ground sate provides direct evidence for the positions of the EPs of the system (Fig. 5(a)), in agreement with our theoretical predictions based on diagonalization of $\hat{\mathcal{H}}_{\text{eff}}$.

A short note regarding the experimental detection of the EPs related to the autoionization of the Helium $2s2p({}^1P)$ AR, is in place at this point. The EP at $( \tilde{\Omega}, \Delta ) = ( \Tilde{\Omega}_{-}, \Delta^s_{-} ) = ( 0.2424Γ, -0.1738Γ )$ lies in a parameter region that is well within the current capabilities of synchrotron sources and seeded Free-electron lasers \cite{ref53,ref54} of short wavelength radiation, sufficient intensity and small bandwidth that can excite the AR.   However, the EP at $( \tilde{\Omega}, \Delta )= ( \tilde{\Omega}_{+}, \Delta^s_{+} ) = ( 8.0265Γ, 5.7538Γ )$ would require a source of high intensity, as it lies in the strong field regime where $\tilde{\Omega} > Γ$ \cite{ref11}. Although the required intensity, which is estimated to be around $1.3 \times 10^{16}$ W/cm$^2$, is available with current Free-electron laser sources, issues such as intensity fluctuations \cite{ref55,ref56} known to affect the excitation of ARs \cite{ref22,ref23,ref24,ref25} and large bandwidth need to be addressed. Their interplay with EPs pose interesting followup studies.

\begin{figure}[t] 
	\centering
	\includegraphics[width=8cm]{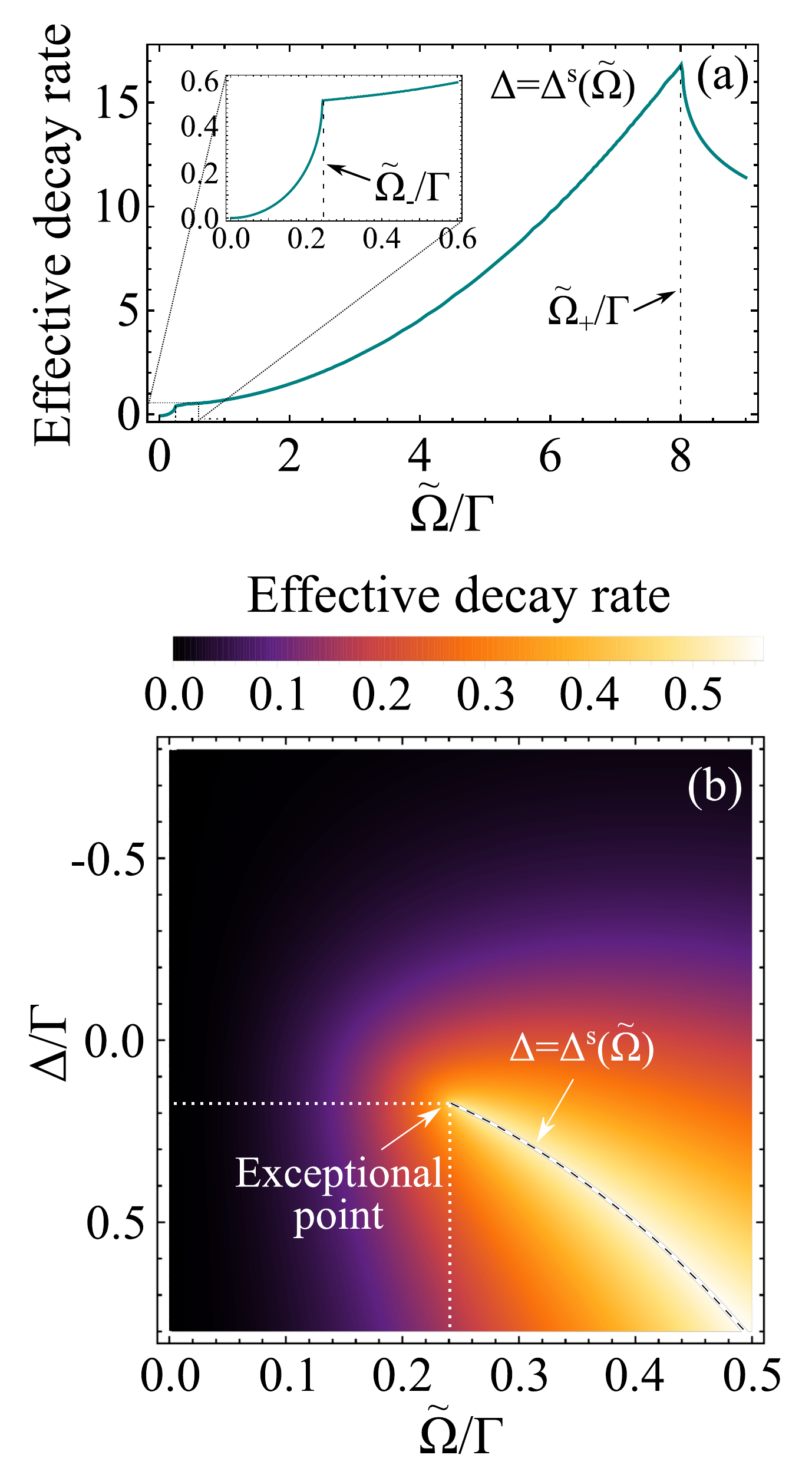}
		\caption[Fig5]{(a) Effective decay rate of the ground state as a function of $\Tilde{\Omega}$ for $q=2.79$ and $\Delta=\Delta^s$. The dashed lines mark the positions of the exceptional points at $\Tilde{\Omega}=\Tilde{\Omega}_{-}=0.2424Γ$ and $\Tilde{\Omega}=\Tilde{\Omega}_{+}=8.0265Γ$. (b) Effective decay rate of the ground state as a function of $\Tilde{\Omega}$ and $\Delta$ for $q=-2.79$. The curved dashed line marks the $\Delta = \Delta^s ( \tilde{\Omega} )$ curve, over which the effective decay rate is maximum. An exceptional point lies at the position $( \tilde{\Omega}, \Delta ) = ( \Tilde{\Omega}_{-}, \Delta^s_{-} ) = ( 0.2424Γ, -0.1738Γ )$.}
		\label{Fig5}
\end{figure}

Finally, in Fig. 5(b) we plot the effective decay rate of the ground state as a function of $\tilde{\Omega}$ and $\Delta$ at the vicinity of the EP that lies in the weak field regime. The effective decay rate maxima lie on the $\Delta=\Delta^s ( \tilde{\Omega} )$ line (curved dashed line) over which the eigenvalues have real radicands. At the tip of this maxima curve we find the weak field EP at the position  $( \tilde{\Omega}, \Delta ) = ( \Tilde{\Omega}_{-}, \Delta^s_{-} ) = ( 0.2424Γ, -0.1738Γ )$ in parameter space. 

\section{Concluding Remarks}
Having unveiled the existence of EPs in single-resonance autoionization and obtained analytical expressions for their positions in terms of the Fano asymmetry parameter, we have further demonstrated their connection with the maxima of the asymmetric ionization profile. Through a quantitative numerical study of the $2s2p({}^1P)$ resonance of Helium, we were led to a reliable method for the observation of EPs, as a function of the parameters of the system, based solely on information about the ionization probability, well within the capabilities of current radiation sources. Our approach and results, based on parameters known from first principles, provide a benchmark for the exploration of EPs in a variety of materials exhibiting a Fano profile, whose potential technological applications are of intense current interest \cite{ref4}. We have moreover prepared the ground for further inquiry on the role of field fluctuations in the observation of EPs in autoionization, as well as questions related to the influence of neighboring ARs, beyond the single-resonance autoionization, which are apt to appear in all systems with a Fano profile. At the same time, the investigation of potentially impactful effects related to phase changes  associated with the encircling of EPs in the parameter space of autoionization, based on the complex topology of the Riemann surfaces in the vicinity of the latter, is a further challenging issue. Overall, our results offer new insights into the interplay between autoionization and non-Hermitian $\mathcal{P} \mathcal{T}$ physics, with connections to several other systems.

\section*{Acknowledgments}
GM would like to acknowledge the Hellenic Foundation for Research and Innovation (HFRI) for financially supporting this work under the 3rd Call for HFRI PhD Fellowships (Fellowship Number: 5525). We are also grateful to D. Kaltsas for useful discussions concerning this work.

\begin{figure}[H] 
	\centering
	\includegraphics[width=3cm]{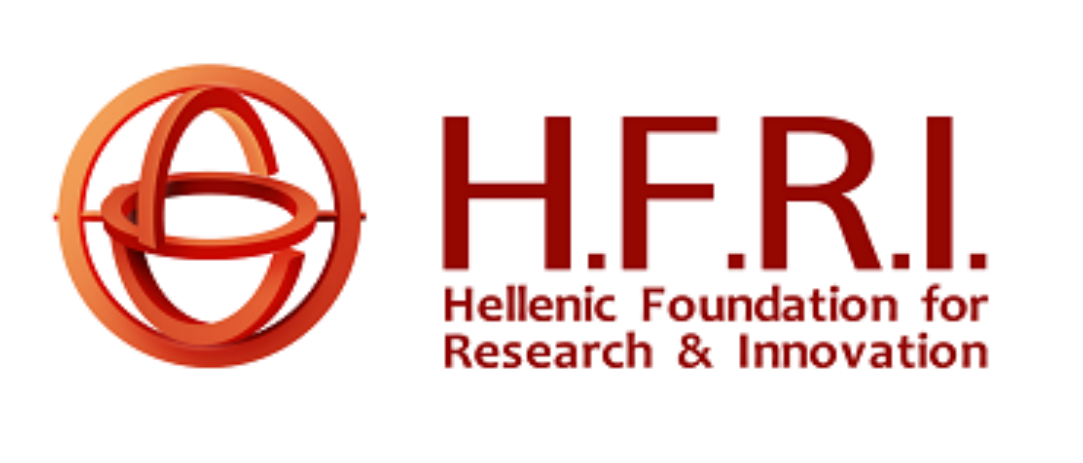}
\end{figure}

\end{document}